# Competing mechanisms of dominant radiative and Auger recombination in hot carrier generation in III-V semiconductor nanowires

Hamidreza Esmaielpour[a], Paul Schmiedeke[a], Nabi Isaev[a], Cem Doganlar[a], Markus Döblinger[b], Jonathan J. Finley[a], Gregor Koblmüller[a]

[a] Walter Schottky Institut, TUM School of Natural Sciences, Technical University of Munich, 85748 Garching, Germany.

[b] Department of Chemistry, Ludwig-Maximilians-University Munich, 81377, Munich, Germany.

**Abstract** – One-dimensional structures such as nanowires (NWs) show great promise in tailoring the rates of hot carrier thermalization in semiconductors with important implications for the design of efficient hot carrier absorbers. However, fabrication of high-quality, phase-pure crystal structures and control of their intrinsic electronic properties can be challenging, raising concerns about the role of competing radiative and non-radiative recombination mechanisms that govern hot carrier effects. Here, we elucidate the impact of crystal purity and altered electronic properties on the hot carrier properties by comparing two classes of III-V semiconductor NW arrays with similar band-gap energies and geometries, yet different crystal quality: one composed of GaAsSb NWs, free of planar stacking defects, and the other InGaAs NWs with a high density of stacking defects. Photoluminescence spectroscopy demonstrates distinct hot carrier effects in both NW arrays; however, the InGaAs NWs with lower crystal quality exhibit stronger hot carrier effects, as evidenced by increased carrier temperature under identical photoabsorptivity. This difference arises from higher rates of Auger recombination in the InGaAs NWs due to their increased n-type conductivity, as confirmed by excitation power-dependent measurements. Our findings suggest that while enhancing material properties is crucial for improving the performance of hot carrier absorbers, optimizing conditions to increase the rates of Auger recombination will further boost the efficiency of these devices.

Thermalization of non-equilibrium hot populations in photovoltaic (PV) solar cells is a significant cause of energy loss in this technology, which limits their maximum conversion efficiency to 33% [1]. Hot carrier solar cells represent one of the third-generation PV devices that aim to enhance efficiency by converting this energy loss into electricity. In a seminal study, Ross and Nozik proposed that this approach could more than double the efficiency of solar cells, provided we can prevent the thermalization of hot carriers through their interactions with phonons and successfully collect them using appropriate energy-selective contacts [2]. Understanding the origin of hot carrier relaxation is, therefore, crucial for designing efficient hot carrier absorbers.

In polar semiconductors, the interactions between hot carriers and longitudinal optical (LO) phonons—known as Fröhlich interactions—are among the primary channels for thermalization [3]. This process involves the transfer of energy from hot carriers to phonons, resulting in a non-equilibrium population of phonons. This phenomenon can slow down the relaxation rates of hot carriers by allowing energy exchange with electrons (or holes), a situation referred to as the phonon-bottleneck effect [4,5,6]. Another mechanism that affects the thermalization rates of hot carriers is intervalley scattering, which occurs quickly following photoabsorption [7,8]. When electrons (or holes) are excited by high-energy photons, they move to satellite valleys before returning to the semiconductor band edge. Although the time that hot carriers spend in these metastable states is relatively short, it effectively extends their lifetime, creating favorable conditions for their collection [9,10]. Recent studies have demonstrated that by utilizing intervalley scattering and extracting hot carriers from the high-energy states, it is possible to increase the open-circuit voltage of PV devices, leading to an improvement in their efficiencies [9,11].

Nanostructured materials, such as quantum wells and one-dimensional (1D) nanowires (NWs), are among the promising candidates for hot carrier absorbers due to their demonstrated phonon-bottleneck effects [12,13,14] and enhanced rates of intervalley scattering [15,16]. These effects arise from the spatial confinement of carriers and the tailored properties of their heterostructures [13,17,18]. Furthermore, this confinement increases the rates of Auger recombination (or Auger heating), leading to stronger hot carrier effects in these materials [19,20]. However, despite their advantages, defects and surface recombination in these compounds can diminish the performance of hot carriers, ultimately reducing the power conversion efficiency of the devices [21]. The impact of defects and the altered electronic properties are a particular concern in 1D-NWs, given the difficulty in obtaining high-quality NWs with phase-pure crystal structure [22] and the associated changes in electronic properties that arise from lattice defects [23]. As such, a complete understanding of the role of defects, and how they impact the hot carrier dynamics in NWs, is still missing.

In this study, we compare the hot carrier effects in two types of NWs that have similar band gap energies and dimensions, particularly in terms of diameter/spatial confinement, yet different crystal purity. One type features a single-crystalline structure, GaAsSb NWs with pure ZB phase, while the other exhibits also ZB-structure but with heavy microstructural disorder, InGaAs NWs. Previous research has shown that the spatial confinement of hot carriers in NWs significantly affects their properties [24]; therefore, comparing these two systems with similar diameters helps us eliminate those variables. Additionally, the absorptivity of the NW arrays under investigation is comparable because their band gap energies, dimensions, and density (the distances between individual NWs in the arrays) align closely. These characteristics make these NW arrays ideal for investigating how crystal quality and trap states influence the dynamics of hot carriers within the system.

The GaAsSb and InGaAs NWs are grown using molecular beam epitaxy (MBE) via vapor-liquid-solid (VLS) and vapor-solid (VS) growth methods, respectively. Detailed information about the growth of these NWs can be found in previous reports [24,25,26]. The GaAsSb NWs consist of an inner core formed through the VLS process, resulting in a composition of GaAs$_{0.75}$Sb$_{0.25}$. The core is surrounded by the spontaneously formed radial shell via a VS growth process, which has a composition of GaAs$_{0.80}$Sb$_{0.20}$ [25]. The results from energy dispersive X-ray (EDX) spectroscopy of the cross-section of the NWs show evidence of two segments of GaAsSb with varying contents of antimony (Sb) and arsenic (As), as illustrated in Figure S1 in the Supplementary Material. A lattice-matched quaternary $In_{0.21}Ga_{0.34}Al_{0.45}As$ shell layer, measuring 15 nm in thickness, is grown around the GaAsSb NWs to passivate them and enhance the quantum yield of luminescence [26]. A scanning electron microscopy (SEM) image of the GaAsSb NWs is displayed in Figure 1(a). These NWs have a diameter of approximately 160 nm and a length of around 6 μm. Figure 1(b) displays a scanning transmission electron microscopy (STEM) image of a single GaAsSb NW, which demonstrates a phase-pure zinc-blende (ZB) crystal structure [25], free of planar defects. In addition, the selected-area electron diffraction (SAED) pattern, as shown in the inset of Figure 1(b), also evidences the absence of planar defects in the ZB structure. This high-quality structure with a low defect density is beneficial as it slows down hot carrier relaxation by reducing the effects of Shockley-Read-Hall (SRH) recombination. Theoretical studies have revealed that when the contribution of SRH recombination increases, hot carriers thermalize at higher rates and lose their energy through phonons [27]. The $In_{0.20}Ga_{0.80}As$ NWs are grown under an entirely catalyst-free, VS-type growth process [24], using the same selective area epitaxial mask pattern and density (pitch of 2 μm) as the GaAsSb NWs. An SEM image of the InGaAs NW array is displayed in Figure 1(c). The diameter and length of the NWs are approximately 160 nm and 3.5 µm, respectively. Similarly, to passivate the dangling bonds on the surface and improve the quantum yield of luminescence, a thin layer (∼ 15 nm) of $In_{0.2}Al_{0.80}As$ is grown around the core NWs [28]. The STEM and SAED results for the InGaAs NWs, shown in Figure 1(d), reveal the presence of a high density of stacking faults and rotational twin defects along the length of the NWs, which developed during growth [29,30].

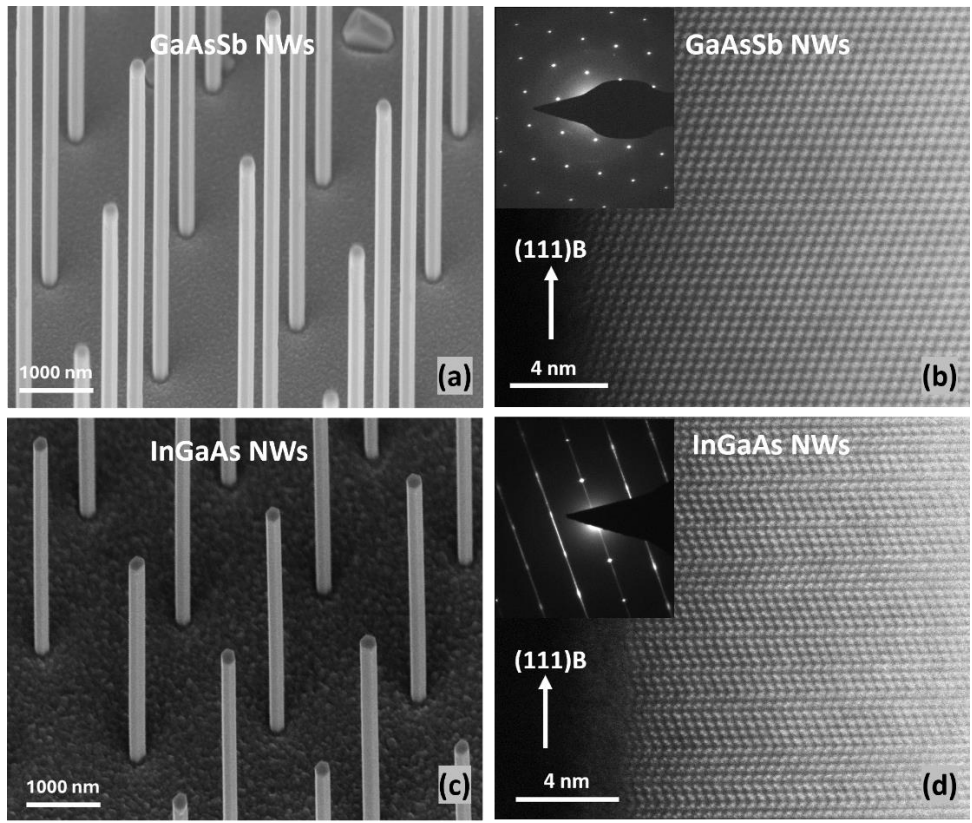

Figure 1. (a,c) SEM and (b,d) STEM images of GaAsSb NWs and InGaAs NWs, respectively. The SAED patterns of the GaAsSb and InGaAs NWs are shown in the insets of (b) and (d). The GaAsSb NWs show evidence of ZB structure without stacking defects. In contrast, the structure of the InGaAs NWs is also based on the ZB structure, however with a high degree of disorder, including stacking faults and rotational twin defects.

The photo-absorptivity of the NW arrays is assessed using finite-difference time-domain (FDTD) simulations conducted with Lumerical® software. The findings indicate that the GaAsSb NW-array yields 88% photo-absorption of incident light under an excitation wavelength of 740 nm, while the InGaAs NW-array has similar photo-absorption of 78%. The investigation into the hot carrier properties of the NW arrays is carried out using micro-photoluminescence (PL) spectroscopy at various excitation power levels and lattice temperatures. The optical setup included a Ti-Sapphire laser with a 740 nm excitation wavelength and a 2 µm spot size, a TRIAX550 imaging spectrograph, and a liquid-nitrogen-cooled Horiba IGA-3000V InGaAs detector to record the PL signal emitted by the samples.

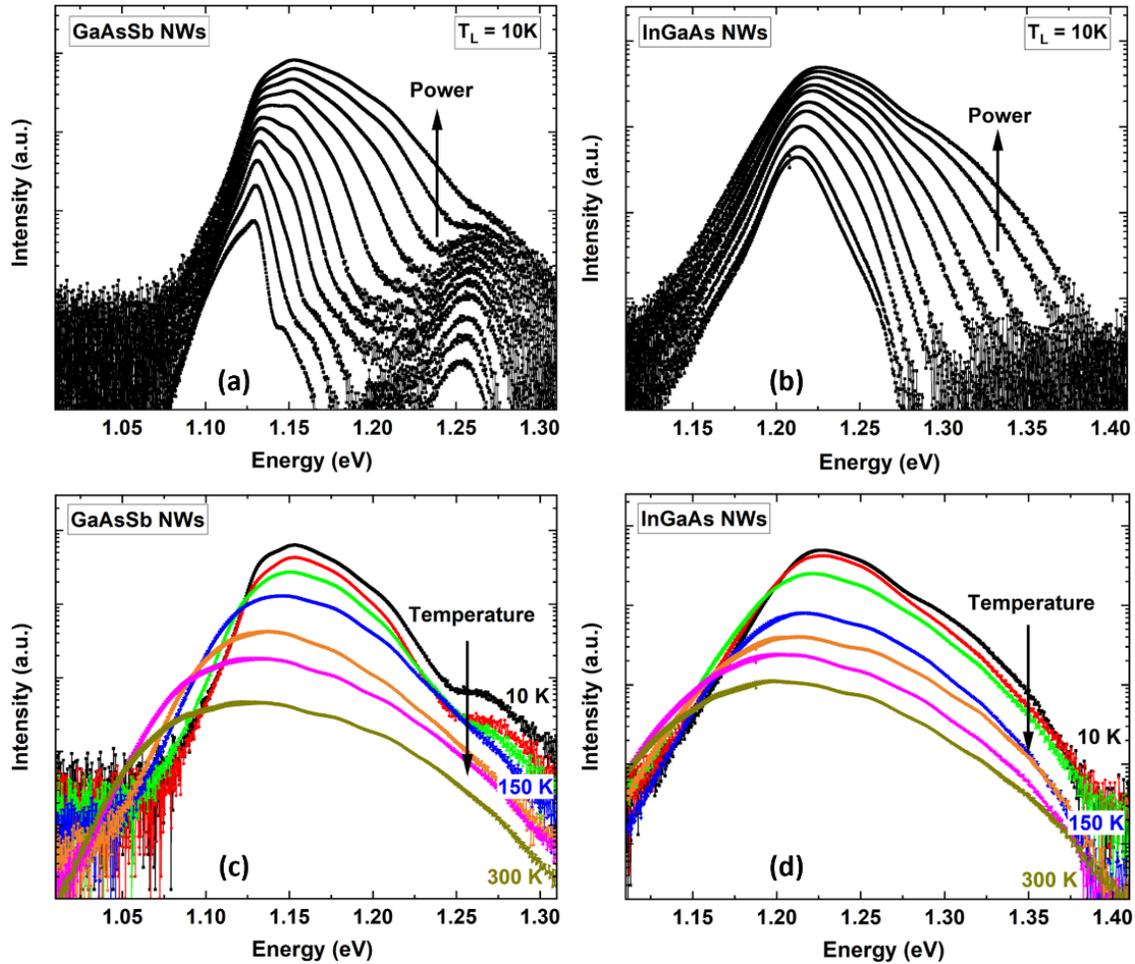

Figure 2. Excitation power-dependent PL spectra at 10 K for (a) the GaAsSb and (b) the InGaAs NW-array. The temperature-dependent PL spectra results for the GaAsSb and InGaAs NWs are shown in (c) and (d), respectively.

Figure 2(a) presents the PL spectra of the GaAsSb NWs at 10 K under different excitation powers. It is observed that as the excitation power increases, the slope on the high-energy side of the PL spectra becomes shallower. This change indicates the generation of hot carriers within the system. At the peak intensity of the PL spectra, two optical transitions are seen: one at 1.13 eV and another at 1.15 eV. These transitions are attributed to the bandgap energies of the inner core, $GaAs_{0.75}Sb_{0.25}$, and the adjacent shell, $GaAs_{0.20}Sb_{0.80}$, surrounding the core [26]. Figure 2(a) further shows that at low excitation powers, most of the photo-generated carriers recombine within the inner GaAsSb core. As the excitation power increases, the band-filling effect allows carriers to transfer into the adjacent GaAsSb shell, which has a larger bandgap energy, resulting in recombination within this segment, as illustrated in Figure 2(a). The optical transition observed at 1.25 eV is attributed to the GaAs NW-stems that are grown beneath the GaAsSb NWs to enhance their growth yield [19]. The GaAs compound experiences tensile strain because it is surrounded by the GaAsSb, causing its emission to shift to such a lower energy value.

The results of the temperature-dependent PL experiments of the GaAsSb NWs are presented in Figure 2(c). As the lattice temperature increases, the peak position of the PL spectra shifts to lower energy values due to lattice expansion. Additionally, at higher lattice temperatures, the optical transition associated with the narrower bandgap of the GaAsSb inner core becomes negligible. This phenomenon occurs because carriers, by receiving thermal energy, move to the larger bandgap $GaAs_{0.80}Sb_{0.20}$ shell, where they recombine. An Arrhenius plot analysis is conducted to determine the activation energies for optical transitions in the NWs, as illustrated in Figure S2 in the Supplementary Material. This analysis identifies two activation energies: $E_1$ = 18 meV and $E_2$ = 78 meV. The first activation energy is close to the difference between the bandgap energies of the VLS- and VS-grown GaAsSb compounds in the NWs, confirming that carriers transfer into the shell at elevated lattice temperatures. The second activation energy is determined by fitting the high-lattice temperature data, which may correspond to a deep-level trap state within the bandgap of the NWs.

In comparison, Figure 2(b) shows the PL spectra of InGaAs NWs at 10 K, observed for the same range of excitation powers. At increased excitation power evidence of hot carrier distribution is seen, similar to the GaAsSb NWs. Figure 2(d) presents the PL spectra of InGaAs NWs at various lattice temperatures, showing the expected redshift with increasing temperature. Since both NW arrays show evidence of hot carrier effects, we proceed with a direct quantitative evaluation and comparison of their carrier temperature and thermalized power density [31,32]. To determine the temperature of hot carriers in these nanostructures, the PL spectra are fitted by the generalized Planck's radiation law as given by [33,34]:

$$I_{PL}(E) = \frac{2\pi\, A(E)\, (E)^2}{h^3 c^2} \left[ exp\left(\frac{E - \Delta\mu}{k_B T}\right) - 1 \right]^{-1}, \qquad (1)$$

where "$I_{PL}$" is the PL intensity, "$E$" the photon energy, "$A$" the absorptivity, "$h$" Planck's constant, "$c$" the speed of light, and "$k_B$" the Boltzmann constant. The temperature and the quasi-Fermi level splitting of photo-generated carriers are represented by "$T$" and "$\Delta\mu$". This relationship can be generalized to estimate the individual temperatures of electrons and holes [35,36]; however, caution is needed during data analysis as the number of fitting parameters increases. Figure 3 illustrates the comparison of the temperature of hot carriers ($\Delta T$: temperature difference between hot carriers and the lattice) versus the absorbed power density of both NW arrays at 10 K and 150 K. For completeness, the hot carrier temperatures at various lattice temperatures and excitation powers for both NW arrays are shown in Figure S3 in the Supplementary Material.

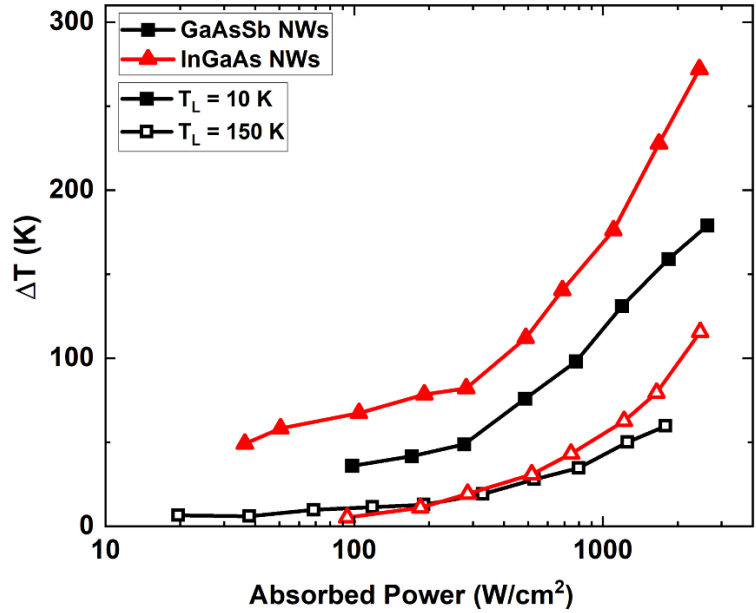

Figure 3. Hot carrier temperature versus the absorbed power density for the GaAsSb NWs (black) and the InGaAs NWs (red) at 10 K (solid symbols) and 150 K (hollow symbols).

As expected, an increase in the temperature of hot carriers is observed in the GaAsSb and InGaAs NW arrays as a function of the absorbed power density ($P_{abs}$). Additionally, Figure 3 shows that, for a given absorbed power density, the temperature of the photo-generated hot carriers in the InGaAs NWs is higher than that in the GaAsSb NWs, even though the crystal quality of the InGaAs NWs is inferior to that of the other structure. Note, both NW arrays are measured using the same excitation power wavelength of 740 nm. However, since the bandgap of these NW structures differs slightly by about ~100 meV, it is more accurate to compare the hot carrier temperature in these structures using the thermalized power density ($P_{th}$), as given by [37]:

$$P_{th} \approx \frac{E_{laser} - E_{gap}}{E_{laser}} P_{abs}, \quad (11)$$

where "$E_{laser}$" and "$E_{gap}$" are the laser and the bandgap energies, respectively. The thermalized power density represents the fraction of absorbed power assigned to hot carriers that are excited beyond the semiconductor's bandgap. This parameter correlates with the excess kinetic energy that hot carriers gain through photogeneration. Figure 4(a) illustrates the results of $\Delta T$ of the NW arrays versus the lattice temperature at a given thermalized power density (450 $W/cm^2$). The NW arrays show a carrier temperature droop at elevated lattice temperatures while the hot carrier temperature of the InGaAs NWs remains higher than the phase-pure GaAsSb NWs.

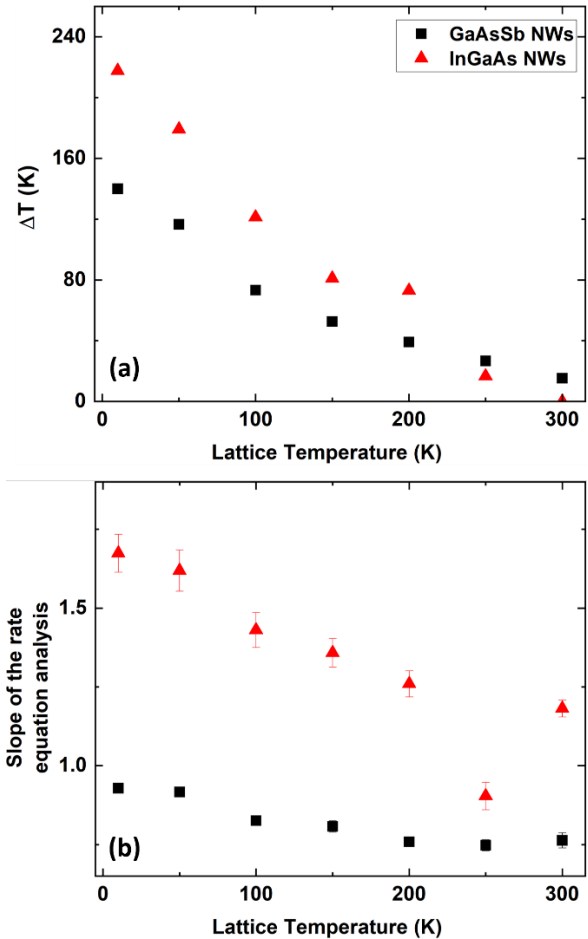

Figure 4. (a) Dependence of $\Delta T$ on the lattice temperature for the GaAsSb NWs (black) and the InGaAs NWs (red) at $P_{th} = 450\ W/cm^2$. (b) Temperature-dependent slope values found from the rate equation analysis for the NW arrays.

Improving crystal quality and minimizing the contribution of non-radiative SRH recombination can enhance the effects of hot carriers [13,41,38], and has previously been found also for InGaAs NWs [22,27]. However, other mechanisms, such as creating phonon-bottleneck effects and increasing Auger recombination rates, can also strengthen hot carrier effects and slow down the relaxation rates of hot carriers [20,28]. Studying the recombination dynamics, using rate equation models, helps us therefore to gain insights into the origin of hot carriers for the two sets of NW-arrays. The rate equation model is defined by [39,40]:

$$P_{Abs.} = A\, I_{PL}^{1/2} + B\, I_{PL} + C\, I_{PL}^{3/2}, \qquad (3)$$

where "$A$", "$B$", and "$C$" are the Shockley-Read-Hall, radiative, and Auger recombination coefficients, respectively. According to this relationship, the dependence of the absorbed power density and the integrated PL intensity determines the dominant recombination mechanism in the system. Figure S4 in the Supplementary shows the results of the natural logarithm of the absorber power density, $ln(P_{Abs.})$, versus the natural logarithm of the integrated PL intensity, $ln(I_{PL})$, at various lattice temperatures. In this analysis, a slope with a magnitude close to 1/2 suggests dominant SRH recombination; a slope of 1 indicates radiative recombination, while 3/2 denotes Auger recombination [40].

Figure 4(b) illustrates the relationship between the slope derived from the rate equation model and the lattice temperature for both types of NW arrays. The GaAsSb NWs display a unique slope across the range of studied power levels (see Supplementary, Fig. S4). In contrast, the recombination dynamics of the InGaAs NWs change significantly from low to high excitation powers. The results presented in Figure 4(b) are obtained at high excitation powers, where the effects of hot carriers become more pronounced. It is observed that GaAsSb NWs with slope values close to unity at low temperatures exhibit recombination dynamics that are similar to radiative recombination. As the lattice temperature increases, these slope values decrease, indicating that there is a rising contribution from SRH recombination at higher temperatures. In contrast, InGaAs NWs display larger slope values (close to 1.5) that resemble Auger recombination across different lattice temperatures. Although the slope derived from the rate equation for InGaAs NWs decreases at elevated lattice temperatures, it remains higher than that of the GaAsSb NWs.

Auger recombination (or Auger heating) is, thus, recognized as an important source of hot carrier generation in our InGaAs NWs, which is much suppressed in the GaAsSb NWs. During the Auger process, one electron-hole pair recombines and transfers its energy to another electron (or hole), promoting it to higher energy states. Consequently, materials with higher rates of Auger heating can exhibit stronger hot carrier effects [20,28,41,42], which aligns well with the observations of higher carrier temperatures in InGaAs NWs compared to GaAsSb NWs. The remaining question that arises from this comparison pertains to the origin of the higher Auger rates in InGaAs NWs, despite the inferior crystalline quality. One reason that we anticipate is a higher (background) free electron concentration in these NWs, which enhances the Auger effect. Indeed, intrinsic InGaAs NWs typically display highly n-type characteristics across a wide range of growth conditions [43,44,45]. The slow growth rate of VS-grown InGaAs NWs, as well as the presence of high stacking defect densities, are known to aggravate unintentional impurity or point defect incorporation [46], which are expected to yield increased n-type carrier concentration. Indeed, recent electrical transport data performed on such VS-grown NWs have shown increased n-type carrier concentrations when stacking defect densities increase [23]. Field-effect transistor measurements performed on the present InGaAs NWs also confirm the highly n-type conductive nature, with estimated free electron concentration in excess of $10^{19}$ cm$^{-3}$. In contrast, in GaAsSb NWs the concentration of free electrons is much reduced due to the formation of Ga$_{Sb}$ antisites [47,48], which pushes the conductivity of GaAsSb NWs towards p-type behavior [49]. As a result, under such a deficiency of electrons, the rates of Auger heating are expected to be reduced significantly. This suggests promising future directions for the control of Auger recombination rates in NWs through intentional doping in order to fabricate hot carrier absorbers that can effectively utilize Auger heating in their structures.

In summary, we investigated the hot carrier effects in NW arrays from two classes of III-V semiconductor materials that crystallize in the native ZB structure, but with quite different structural quality: one composed of GaAsSb NWs, free of planar stacking defects, and the other made up of highly n-type conductive InGaAs NWs with a high-density of stacking defects. Otherwise, the NW arrays exhibited similar absorptivity, dimensions, and band-gap, making them suitable candidates for studying how crystal quality and intrinsic electronic properties influence the properties of hot carriers and their recombination dynamics. Both NW arrays show evidence of hot carrier populations; however, the effects of hot carriers in the InGaAs NWs are more pronounced at a given absorbed power density compared to the GaAsSb NWs. While it is generally expected that higher-quality crystal structures with fewer scattering centers would enhance the generation of hot carriers and decrease their thermalization rates, the InGaAs NWs exhibit higher temperatures than the phase-pure GaAsSb NWs. The origin of this effect is attributed to increased rates of Auger recombination in the InGaAs NWs, as indicated by the rate equation model, which suggests a significant heating pathway for carriers due to their high n-type conductivity. This study highlights that although high crystal quality and low defect density are important for designing efficient hot carrier absorbers, other competing mechanisms that can generate robust hot carrier effects must also be considered.

This work was supported by the European Research Council (ERC project QUANtIC, ID: 771747) and the Deutsche Forschungsgemeinschaft (German Research Foundation (DFG)) under Germany's Excellence Strategy via the Cluster of Excellence e-conversion (EXC 2089/1-390776260)). The authors further acknowledge support by the Marie Sklodowska Curie Action (MSCA) via the TUM EuroTechPostdoc2 Grant Agreement (No. 899987). The authors further thank H. Riedl for support with the MBE system.

**Data Availability**

The data that support the findings of this study are available from the corresponding author upon reasonable request.

# Competing mechanisms of dominant radiative and Auger recombination in hot carrier generation in III-V semiconductor nanowires

Hamidreza Esmaielpour[a], Paul Schmiedeke[a], Nabi Isaev[a], Cem Doganlar[a], Markus Döblinger[b], Jonathan J. Finley[a], Gregor Koblmüller[a]

[a] Walter Schottky Institut, TUM School of Natural Sciences, Technical University of Munich, 85748 Garching, Germany.

[b] Department of Chemistry, Ludwig-Maximilians-University Munich, 81377, Munich, Germany.

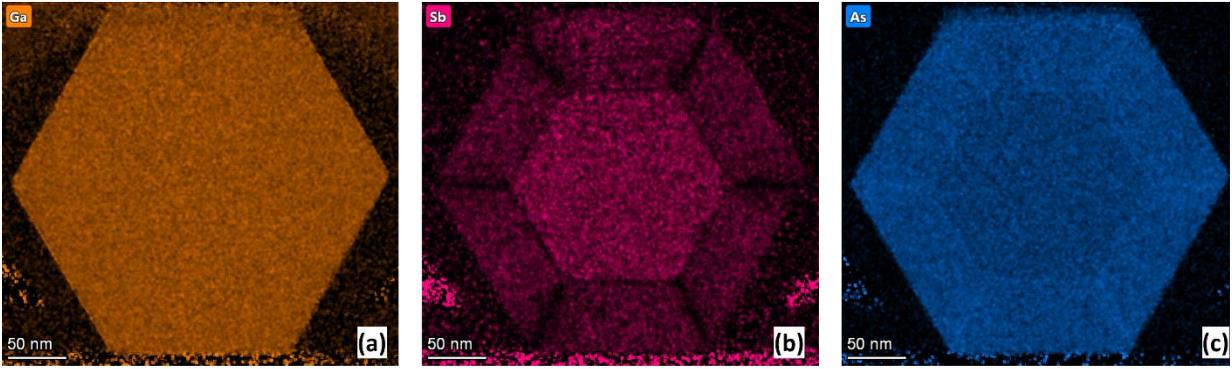

Figure S1: Results of scanning transmission electron microscopy (STEM) analysis on the cross-section of the GaAsSb nanowires (NWs) for the elements (a) Ga, (b) Sb, and (c) As. It is observed that the concentration of Sb is higher in the core formed by the vapor-liquid-solid (VLS) method compared to the vapor-solid (VS) grown shell. In contrast, the As content shows the opposite trend in the cross-section.

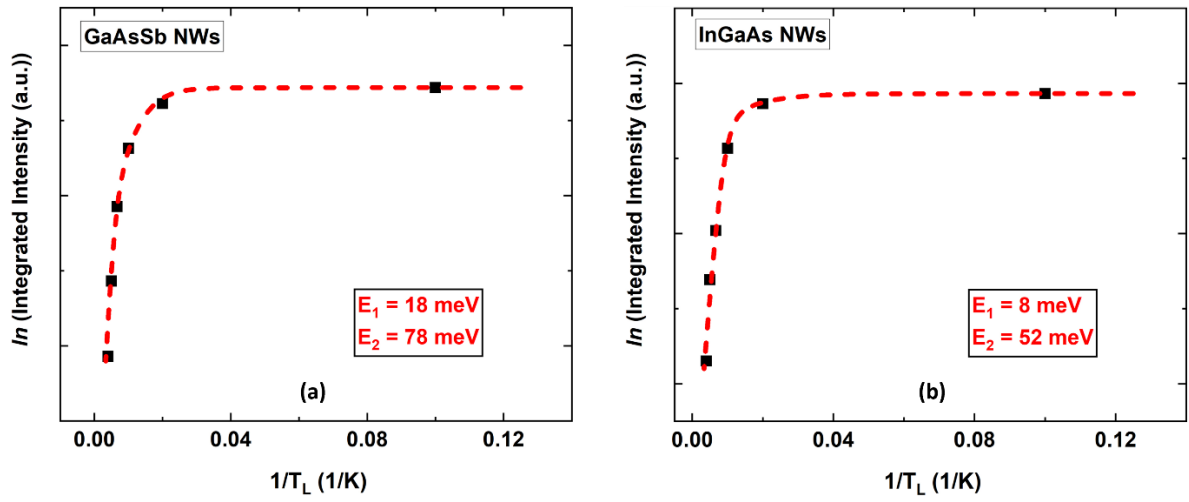

Figure S2: The Arrhenius plot for the temperature-dependent PL spectra of (a) the GaAsSb NWs and (b) the InGaAs NWs.

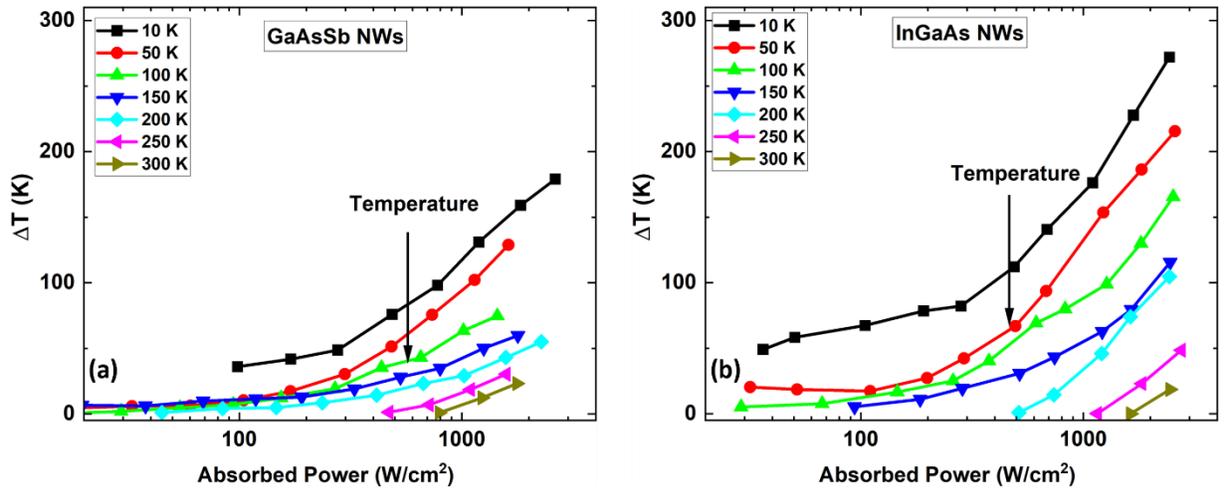

Figure S3: Hot carrier temperature ($\Delta T$) as a function of the absorbed power density for (a) the GaAsSb NWs, and (b) the InGaAs NWs at various lattice temperatures.

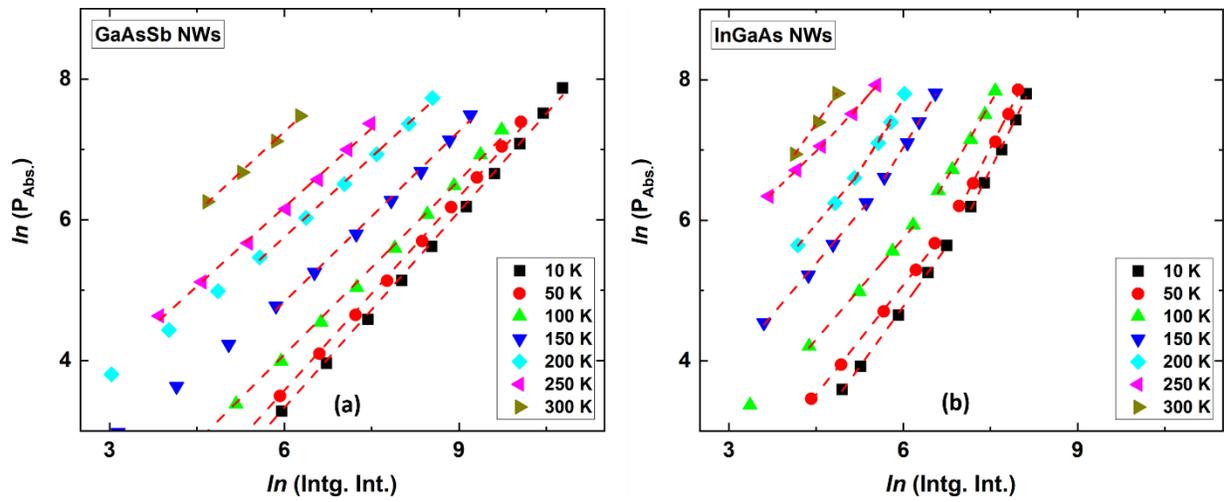

Figure S4: Results of the rate equation model for (a) the GaAsSb NWs and (b) the InGaAs NWs at various lattice temperatures. The GaAsSb NWs exhibit a single unique slope, while the InGaAs NWs display two distinct slope values. This indicates the presence of different recombination dynamics within their structures across the studied power range.